\documentclass[aps,prd,reprint,nofootinbib,superscriptaddress]{revtex4-2}

\usepackage{amsmath,amssymb,bm}
\usepackage{graphicx}
\usepackage{booktabs}
\usepackage{hyperref}
\usepackage[dvipsnames]{xcolor}

\newcommand{\jim}{\textsc{Jim}}
\newcommand{\jimlensing}{\textsc{Jim-lensing}}

\newcommand{\dd}{\mathrm{d}}

\newcommand{\commentout}[1]{}

\usepackage{lineno}
\usepackage{orcidlink}

\begin{document}

\title{GPU-accelerated Bayesian inference for diffracted gravitational wave signals}

\author{Zihao Lin\,\orcidlink{https://orcid.org/0009-0002-1763-4408}}
\email{191850100@smail.nju.edu.cn}
\affiliation{School of Astronomy and Space Science, Nanjing University, Nanjing 210023, China}
\affiliation{Department of Physics, Faculty of Arts and Sciences, Beijing Normal University, Zhuhai 519087, China}

\author{Jef Heynen\,\orcidlink{https://orcid.org/0009-0007-5496-2159}}
\email{jef.heynen@uclouvain.be}
\affiliation{Centre for Cosmology, Particle Physics and Phenomenology (CP3), Universit\'e catholique de Louvain, B-1348 Louvain-la-Neuve, Belgium}

\author{Mick Wright\,\orcidlink{https://orcid.org/0000-0003-1829-7482}}
\affiliation{Institute for Gravitational and Subatomic Physics (GRASP), Department of Physics, Utrecht University, Princetonplein 1, 3584 CC Utrecht, The Netherlands}
\affiliation{Nikhef---National Institute for Subatomic Physics, Science Park 105, 1098 XG Amsterdam, The Netherlands}

\author{Thomas C. K. Ng\,\orcidlink{https://orcid.org/0000-0002-9491-1598}}
\affiliation{Institute for Gravitational and Subatomic Physics (GRASP), Department of Physics, Utrecht University, Princetonplein 1, 3584 CC Utrecht, The Netherlands}
\affiliation{Nikhef---National Institute for Subatomic Physics, Science Park 105, 1098 XG Amsterdam, The Netherlands}

\author{Chris Van Den Broeck\,,\orcidlink{https://orcid.org/0000-0001-6800-4006}}
\affiliation{Institute for Gravitational and Subatomic Physics (GRASP), Department of Physics, Utrecht University, Princetonplein 1, 3584 CC Utrecht, The Netherlands}
\affiliation{Nikhef---National Institute for Subatomic Physics, Science Park 105, 1098 XG Amsterdam, The Netherlands}

\author{Justin Janquart\,\orcidlink{https://orcid.org/0000-0003-2888-7152}}
\affiliation{Centre for Cosmology, Particle Physics and Phenomenology (CP3), Universit\'e catholique de Louvain, B-1348 Louvain-la-Neuve, Belgium}
\affiliation{Royal Observatory of Belgium, Avenue Circulaire, 3, 1180 Uccle, Belgium}

\begin{abstract} 
  As the rate of gravitational wave detections continues to grow, the probability of observing a
  gravitationally lensed gravitational wave also increases. When the
  Schwarzschild radius of the lens is comparable in size to the gravitational wave wavelength, wave-optics effects induce a frequency-dependent
  modulation on the signal. CPU-based parameter estimation analyses for
  wave-optics lensing are computationally intensive, limiting the scale of
  studies to determine the significance of any potential lensing candidate.  In
  this work, we implement a point-mass lens model in \jim, a high-performance
  GPU-based parameter estimation pipeline. This enables, for the first time, GPU-accelerated Bayesian inference in full-wave optics, including the computation of Bayes factors, without the need for training data. We call this framework \textsc{Jim-lensing}. Using this framework, we are able to speed up a full parameter estimation run under the point-mass lens hypothesis from more than 10 hours to the order of 10 minutes compared to the current state-of-the-art CPU-based pipeline.
  We show that our implementation
  is stable and obtains comparable results to this state-of-the-art pipeline. The substantial speed gain over existing CPU pipelines
  paves the way for analyzing gravitational wave data under a broader range of
  lensing models with greater capacity to produce robust frequentist
  significance estimates.
\end{abstract}


\maketitle

\section{Introduction}

When a gravitational wave (GW) passes near a massive object, the curvature of
spacetime caused by the intervening mass can modify the propagation of that
GW~\citep{Deguchi:1986,Schneider:1992bmb,Wang:1996as,Nakamura:1997sw,Takahashi:2003ix}.
This gravitational lensing effect is well studied for electromagnetic waves but
has not yet been detected for
GWs~\citep{Hannuksela:2019kle,LIGOScientific:2021izm,LIGOScientific:2023bwz,Janquart:2023mvf,LIGOScientific:2025cwb}.
For the heaviest lenses, the lensing effect is described by geometric optics,
in which multiple images of the same original GW are produced, each with a
different arrival time, magnification and phase
shift~\citep{Wang:1996as,Takahashi:2003ix}. However, for lighter lenses the
wavelength of the GW becomes comparable to the Schwarzschild radius of the
lens, and wave-optics effects which impart frequency-dependent modulations of
the signal~\citep{Deguchi:1986,Nakamura:1997sw,Takahashi:2003ix} become
observable. These lensing effects are modeled via an amplification factor,
which is analytically solvable only for a few simple lens models, such as the
point mass lens and the singular isothermal sphere
(SIS)~\citep{Takahashi:2003ix,Matsunaga:2006uc}. For more complex lens models,
the amplification factor must be found by numerically solving the diffraction
integral~\citep{Ulmer:1994ij,Diego:2019lcd,Lai:2018rto,Cheung:2020okf,Yeung:2021chy,Cheung:2024ugg,Mishra:2021xzz,Villarrubia-Rojo:2024xcj}.
Once observed, lensed GWs can be used to probe a wider array of scientific
avenues to deepen our understanding of the universe. Examples of these avenues
include studying smaller-scale galaxy
content~\citep{Mishra:2021xzz,Cheung:2024ugg}, probing the nature of dark
matter~\citep{Urrutia:2021qak,Tambalo:2022wlm} and providing alternative tests
of general
relativity~\citep{Ezquiaga:2020dao,Goyal:2020bkm,Goyal:2023uvm,Narola:2023viz}.

Recent works have integrated amplification factors for different lens profiles
in Bayesian inference pipelines in order to perform parameter estimation (PE)
and model selection on GW signals under the microlensing
hypothesis~\citep{Wright:2021cbn, LIGOScientific:2023bwz,
LIGOScientific:2025cwb, Shan:2025dcd, Goyal:2025eqo}. However, they are based
on standard PE frameworks, such as \textsc{Bilby}~\citep{Ashton:2018jfp},
\textsc{LALInference}~\citep{Veitch:2014wba}, or
\textsc{PyCBC}~\citep{Biwer:2018osg} which employ CPU implementations of nested
sampling~\citep{Skilling, Skilling:2006gxv} and Markov chain Monte Carlo
(MCMC)~\citep{Tierney:1994}, making them computationally expensive. This introduces
an operational challenge with the ever increasing rate of GW detections
expected in future observing runs~\citep{KAGRA:2013rdx} and especially as
third-generation detectors, such as the Einstein Telescope~\citep{Hild:2010id,
Punturo:2010zz, Branchesi:2023mws} and Cosmic Explorer~\citep{Reitze:2019iox},
come online.  This computational cost also limits the scale of backgrounds of
simulated signals to be analysed by these standard analyses. These backgrounds
are used to estimate the statistical significance of any potentially
microlensed GW events. This is most exemplified by
GW231123~\citep{LIGOScientific:2025rsn}, which showed support for microlensing,
but where computational and temporal constraints limited the LIGO--Virgo--KAGRA
(LVK) collaboration to a background of only 254 events, leading to an upper
bound on the $p$-value, including the trials factor, of only
$28\%$~\citep{LIGOScientific:2025cwb,Goyal:2025eqo}.

To address these challenges, GPU-based PE pipelines have been developed to
accelerate the inference process. Some of these have also included machine
learning aspects such as \textsc{Dingo}~\citep{Green:2020dnx,Green:2020hst,Dax:2021tsq}, which makes use of
neural posterior estimation---training a neural network to map input data to a posterior distribution
enabling, once trained, the execution of PE in seconds. Recent efforts have
targeted the lensing analysis problem specifically with the development of
\textsc{DINGO-Lensing}~\citep{Chan:2025pdf, Chan:2025kyu, Caldarola:2025oxr}, obtaining accurate PE results and showing that large-scale backgrounding is possible with such methods. However,
a challenge to these methods is the necessity
of training on particular simulated scenarios, 
which must be done for each potential data setup and sensitivity, and which can be violated during real data analysis, 
leading to out-of-training data. 
This is exemplified in~\citet{Caldarola:2025oxr},
where the authors found that the sampling efficiency dropped when analyzing lensed events with a network trained on unlensed simulations.

An alternative GPU-based approach to this problem is \jim~\citep{Wong:2023lgb},
which exploits the \textsc{JAX}-based~\citep{jax2018github} automatically
differentiable gravitational waveform library
\textsc{ripple}~\citep{Edwards:2023sak} to achieve GPU-accelerated likelihood
calls when combined with a \textsc{JAX}-based sampler such as
\textsc{BlackJAX}~\citep{cabezas:2024}. Recently, a sampler~\citep{Prathaban:2025qgg} has been developed, based on \textsc{BlackJAX}, that incorporates sampling methods comparable to
those used with the well-established~\textsc{dynesty}~\citep{Speagle_2020}
sampler used for GW analyses, however, accomplishing runs in a matter of
minutes. Whilst this is individually slower than the above approach, it does
not depend on an initial training set meaning that it will not suffer from out-of-training data and that it may be more flexibly applied.

In this work, we implement the isolated point mass lensing amplification factor
in the wave-optics regime in a \textsc{JAX}-compatible manner enabling Bayesian
inference on these lens diffracted GW signals at comparable accuracy to
classical methods while retaining the GPU-acceleration and without training
data concerns. Although astrophysical point-mass lenses are expected to be rare~\cite{Vujeva:2026xfh}, they are one of the very few analytically solvable lens models and they approximate reasonably well other spherically symmetric lens models. Furthermore, the point mass is a standard lens model used in LVK analyses~\citep{LIGOScientific:2023bwz, LIGOScientific:2025cwb}, allowing for direct comparison with existing pipelines. Our methodology is available in the package,
\jimlensing{}, built atop \jim. We demonstrate that the framework passes
the percentile-percentile (PP) test and obtains compatible posterior
distributions on both source and lens parameters with the state-of-the-art
\textsc{Gravelamps}~\citep{Wright:2021cbn}, commonly used in LVK
analyses~\citep{LIGOScientific:2023bwz, LIGOScientific:2025cwb}.

The paper is structured as follows: In Section \ref{sec:prelim}, we summarize the relevant GW lensing theory and describe the \jim{} framework. In Section \ref{sec:jimlensing}, we present the implementation of \jimlensing{}. In Section \ref{sec:results}, we validate the framework through waveform mismatch and percentile-percentile tests, compare its PE results with \textsc{Gravelamps}, and investigate the recovery of Bayes factors. We summarize our conclusions and discuss future prospects in Section \ref{sec:conclusion}.

\section{Preliminaries}
\label{sec:prelim}
\subsection{Bayesian inference}

Analysis of GW signals is performed within a Bayesian framework. Estimates of the parameters, $\bm{\theta}$, of a model $\mathcal{H}$, from
data, $d$, are characterised by their posterior distribution, $p(\bm{\theta} |
d,\mathcal{H})$, which may be given by Bayes' theorem:
\begin{equation}
  p(\bm{\theta}|d,\mathcal{H}) = \frac{\mathcal{L}(d|\bm{\theta},\mathcal{H})\,\pi(\bm{\theta}|\mathcal{H})}{\mathcal{Z}_\mathcal{H}},
    \label{eq:bayes_theorem}
\end{equation}
where $\mathcal{L}(d|\bm{\theta},\mathcal{H})$ is the likelihood, $\pi(\bm{\theta}|\mathcal{H})$ is the
prior on the parameters, and $\mathcal{Z}_\mathcal{H} = \int \mathrm{d}\bm{\theta}\,
\mathcal{L}(d|\bm{\theta},\mathcal{H})\pi(\bm{\theta}|\mathcal{H})$ is the evidence, which normalises
the posterior. Assuming stationary, Gaussian noise, the likelihood takes the
form~\citep{Veitch:2014wba}
\begin{equation}
    \ln \mathcal{L}(d|\bm{\theta},\mathcal{H}) = -\frac{1}{2}\left\langle d - h(\bm{\theta})\,,\,d - h(\bm{\theta})\right\rangle + \text{const.},
    \label{eq:gw_likelihood}
\end{equation}
where $h(\bm{\theta})$ is the waveform model evaluated at parameters
$\bm{\theta}$, and $\left\langle a\,,\,b\right\rangle$ denotes the
noise-weighted inner product defined by
\begin{equation}
    \langle a\,,\,b\rangle = 4\text{Re}\int_{f_\text{min}}^{f_\text{max}}\frac{\tilde{a}(f)\tilde{b}^*(f)}{S_n(f)}\mathrm{d}f\,,
\end{equation}
where $S_n(f)$ is the one-sided power spectral density (PSD) characterizing the
noise. Bayesian hypothesis testing between two models, $\mathcal{H}_1$ and $\mathcal{H}_2$, is performed by computing the posterior odds:
\begin{equation}
    \frac{p(\mathcal{H}_2|d)}{p(\mathcal{H}_1|d)} = \frac{p(\mathcal{H}_2)}{p(\mathcal{H}_1)}\frac{p(d|\mathcal{H}_2)}{p(d|\mathcal{H}_1)} = \frac{p(\mathcal{H}_2)}{p(\mathcal{H}_1)}\frac{\mathcal{Z}_{\mathcal{H}_2}}{\mathcal{Z}_{\mathcal{H}_1}}\,,
\end{equation}
where the first factor on the right-hand side is called the prior odds and the second factor is called the Bayes factor, which is the ratio of the evidences and is commonly denoted as $\mathcal{B}^{\mathcal{H}_2}_{\mathcal{H}_1}$.

\subsection{Jim}

\textsc{Jim}~\citep{Wong:2023lgb,Wouters:2024oxj} is a free, open-source, software package
designed to perform Bayesian inference on GW
signals rapidly and accurately. It leverages the \textsc{JAX}~\citep{jax2018github} python library which enables high speed array computation through access to GPU-based calculations to significantly speed up the inference process compared to CPU-based calculations. \textsc{Jim} is
modular, allowing the usage of a number of different sampling methods and
techniques. In this work, we use the \textsc{BlackJAX-ns}~\citep{cabezas:2024}
implementation of nested sampling, and in particular make use of the
`acceptance-walk' sampling method implementation developed
by~\citet{Prathaban:2025qgg}. This yields an implementation mirroring the
\textsc{dynesty}~\citep{Speagle_2020} implementation used in current
state-of-the-art PE frameworks e.g the \textsc{Bilby} analyses of LVK
detections~\citep{LIGOScientific:2026ifv}

\subsection{Lensed gravitational waveform}
In what follows, we use geomtric units: $c=G=1$.
Using the thin-lens approximation, the lensed waveform $\tilde{h}_L(f)$ and the
unlensed waveform $\tilde{h}_U(f)$ are related through the amplification factor
$\tilde{F}(f)$ as~\citep{Takahashi:2003ix}
\begin{equation}
    \tilde{h}_L(f) = \tilde{F}(f)\tilde{h}_U(f)\,.
    \label{eq:relation_lensed_unlensed_wf}
\end{equation}

Using an arbitrary normalisation length, $\xi_0$, we scale the lens plane
coordinates by $\xi_0$ to obtain dimensionless lens plane coordinates $\bm{x}$,
and we scale the source position in the source plane by $\xi_0D_\text{OS} /
D_\text{OL}$, to obtain the dimensionless source position $\bm{y}$, where
$D_\text{OL}$ and $D_\text{OS}$ are the angular distances between the observer
and the lens and source, respectively~\citep{Schneider:1992bmb}. The
amplification factor is then given as the following diffraction
integral~\citep{Takahashi:2003ix}:
\begin{equation}
    \tilde{F}(w)
    =
    \frac{w}{2\pi i}
    \int \mathrm{d}^{2}\bm{x}\,
    \exp\!\left[iwT(\bm{x},\bm{y})\right],
    \label{eq:diffraction_integral}
\end{equation}
where, for a redshifted lens mass $M_L$, $w = 8\pi M_L f$ is the dimensionless
frequency, and $T(\bm{x},\bm{y})$ is the dimensionless Fermat potential, whose
form depends on the gravitational potential of the lens. Note that the
amplification factor only depends on the lens mass through the dimensionless
frequency. 

For a point-mass lens, the dimensionless Fermat potential
is~\citep{Takahashi:2003ix}
\begin{equation}
    T(x,y)
    =
    \frac{1}{2}
    \left|x-y\right|^{2}
    -
    \ln x
    -
    \phi_{\rm m}(y)\,,
\end{equation}
where $y$ is the dimensionless impact parameter of the source. The final term ensures the minimum of the Fermat
potential to be zero. For a point-mass lens,
Eq.~\eqref{eq:diffraction_integral} admits an analytic
expression~\citep{Takahashi:2003ix}
\begin{equation}
\begin{split}
  \tilde{F}(w)={}&
  \exp\!\left[
  \frac{\pi w}{4}
  +\frac{iw}{2}
  \left\{
  \ln\!\left(\frac{w}{2}\right)-2\phi_{\rm m}(y)
  \right\}
  \right]
  \\
  &\times
  \Gamma\!\left(1-\frac{iw}{2}\right)
  \\
  &\times
  {}_{1}F_{1}\!\left(
  \frac{iw}{2},1;\frac{iwy^{2}}{2}
  \right).
\end{split}
\label{eq:amp_factor_point_lens}
\end{equation}
Here, $\Gamma$ denotes the Gamma function, ${}_{1}F_{1}$ is the confluent
hypergeometric function of the first kind, and $\phi_m(y)$ is given by $(x_m -
y)^2/2-\ln x_m$ where $x_m = (y+\sqrt{y^2+4})/2$. In the GO approximation, the
diffraction integral is dominated by the stationary points of the Fermat
potential. The amplification factor is then given by 
\begin{equation}
    \tilde{F}_\text{GO}(w) = |\mu_+|^{1/2} - i|\mu_-|^{1/2}e^{iw\Delta T}\,,
    \label{eq:amp_factor_GO}
\end{equation}
where the magnifications of the images are given by
\begin{equation}
\mu_{\pm} = \frac12 \pm
\frac{y^2+2}{2y\sqrt{y^2+4}}
\end{equation}
and the time delay between them is 
\begin{equation}
\Delta T =
\frac{y\sqrt{y^2+4}}{2}+\ln\frac{\sqrt{y^2+4}+y}{\sqrt{y^2+4}-y}.
\end{equation}
Note that the
point-mass lens introduces two additional variables to our inference problem,
the source position $y$ and the redshifted lens mass $M_L$.

\section{Jim-lensing}
\label{sec:jimlensing}
\subsection{Jim-lensing}
We implement a point-mass lensed waveform generator in \textsc{Jim}. This
generator takes as its base the efficient
\textsc{ripple}~\citep{Edwards:2023sak} waveforms from the binary parameters.
It then generates the amplification factor from the lens parameters and
computes the total lensed waveform via
Eq.\eqref{eq:relation_lensed_unlensed_wf}. In this work, we use
the \textsc{ripple} implementation of the aligned-spin
\textsc{IMRPhenomD}~\citep{Husa:2015iqa, Khan:2015jqa} model, although the method is directly applicable to other \textsc{ripple} waveforms. For the amplification factor, similarly to the
\textsc{Gravelamps} framework~\citep{Wright:2021cbn}, we first pre-compute a grid
of amplification factor values which is used as a basis for interpolation for
actual waveform calls.

Conversion of the amplification factor function calls to \textsc{JAX} in this
work, requiring an implementation of the Gamma and confluent hypergeometric
functions that can tolerate complex arguments. Recently, the implementation for the Gamma function has been updated in the \texttt{jax.scipy.special} library~\citep{jax2018github} to allow for complex inputs, making use of the Lanczos approximation~\citep{Lanczos:1964}.
However, this is not the case for the confluent hypergeometric function. To address this, we implement a \textsc{JAX} compatible computation of this function. 

\begin{figure}[t]
    \includegraphics[width=\linewidth]{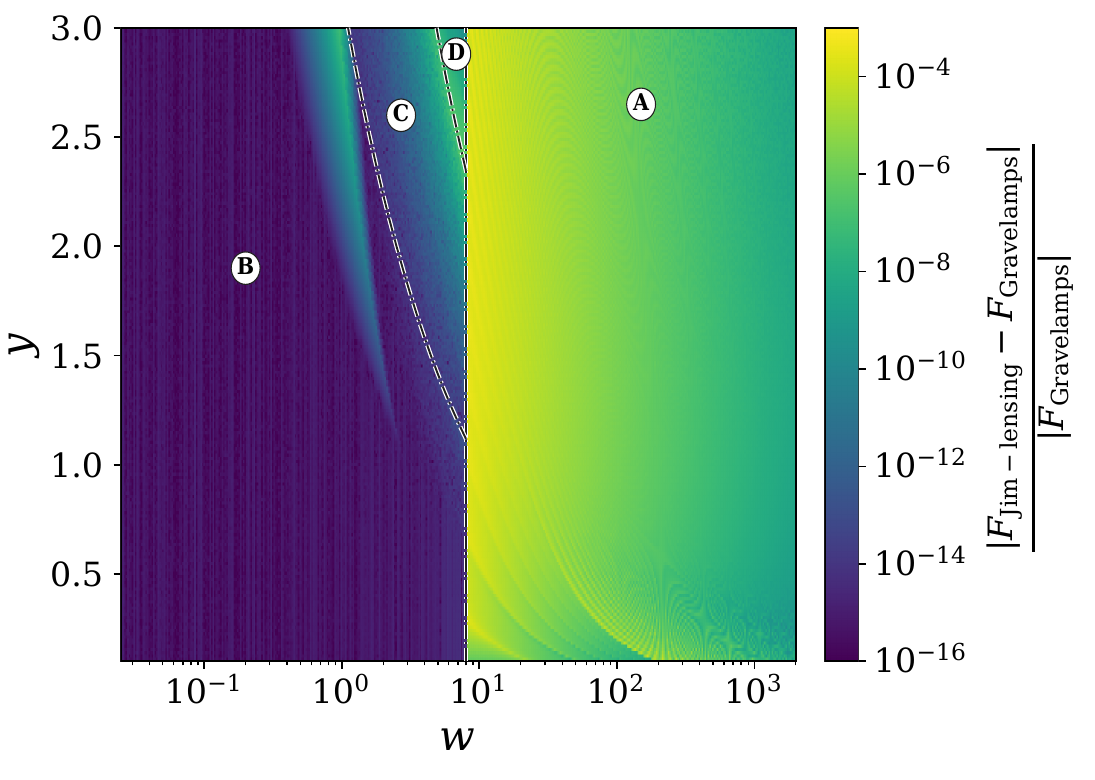}
    \caption{Relative difference between amplification factor values computed in \textsc{Gravelamps} and \jimlensing{} for different lens parameters. The definitions of the different regions can be found in Table \ref{tab:1F1_branches}}
    \label{fig:amp_factor_err}
\end{figure}

We adopt the numerical strategy used
by~\citet{Villarrubia-Rojo:2024xcj} in which the $(w, y)$-space is separated
into four differing regions based on the values of
\begin{equation}
    u = \frac{w}{2}\,,\qquad \zeta = \frac{wy^2}{2}\,,
\end{equation}
and apply in each region a different approximation method to the confluent
hypergeometric function. The definitions of the regions and which numerical
approximation applies where can be found in Table
\ref{tab:1F1_branches}. The details of the different approximations are given in Appendix~\ref{subsec:1F1}.
Using this strategy, the full grid is computed in less than a minute and as may
be seen from Fig.~\ref{fig:amp_factor_err}, the relative error compared to the
full calculation of the amplification factor within \textsc{Gravelamps} is
below $10^{-5}$ for the bulk of the $(w,y)$-space, however, we do note that
there is a jump in the relative error on the switch to the Temme
expansion. This gives rise to a region with relative errors from $10^{-5}$ to $10^{-3}$. However, this region only comprises a small vertical strip in $(w,y)$-space and as may be seen in
Section~\ref{subsec:mismatch}, this does not lead to a measurable difference in
the final waveforms.

\begin{table}[t]
\caption{Different regions and approximations of the confluent hypergeometric function.}
\label{tab:1F1_branches}
\begin{ruledtabular}
\begin{tabular}{clc}
& Region & Approximation \\
\hline
A & $u>4$ & Temme expansion \\
B & $u\leq4$, $\zeta<5$ & Direct power series \\
C & $u\leq4$, $5\leq\zeta\leq22$ & Downward recurrence \\
D & $u\leq4$, $\zeta>22$ & Large-$\zeta$ asymptotic expansion \\
\end{tabular}
\end{ruledtabular}
\end{table}

\subsection{Sampling configuration} 

We perform parameter estimation with \textsc{Jim-Lensing} using
the \textsc{BlackJAX-ns} sampler integrated within \textsc{Jim} making use of
the `acceptance-walk' implementation specifically~\citep{cabezas:2024,
Prathaban:2025qgg}. The sampler is configured with $1000$ live points, an expected number of accepted proposal steps of $60$, a live points deletion fraction of $0.5$, a maximum number of proposal steps of $5000$ and a termination criterion of the estimated increase in log-evidence of the live points less than $0.1$. Furthermore, we approximate the likelihood by marginalising over phase~\citep{Romero-Shaw:2020owr}, which is a technique used to speed up the sampling process and which is valid for waveforms that only include the dominant mode, which is the case for our chosen \textsc{IMRPhenomD} waveform model~\citep{Husa:2015iqa,Khan:2015jqa}. For the comparison with \textsc{Gravelamps}, we use the \textsc{dynesty} sampler~\citep{Speagle_2020} configured with $1000$ live points---noting that despite
similarities in terminology, as noted in~\citet{Prathaban:2025qgg} these are not
precisely equivalent numbers---, an expected number of accepted proposal steps of $60$, a maximum number of proposal steps of $5000$ and a termination criterion of the estimated increase in log-evidence of the live points less than $0.1$. We do not use phase marginalisation. The above settings are consistent with those used in the LVK~\citep{LIGOScientific:2025cwb}.

\section{Results}
\label{sec:results}
\subsection{Mismatch study} \label{subsec:mismatch}

We first validate our implementation by computing the mismatch between the
lensed waveforms generated by \textsc{Gravelamps} and \textsc{Jim-lensing}. The
mismatch between two waveforms $h_1$ and $h_2$ is defined as

\begin{equation}
    \mathcal{M}(h_1,h_2) = 1 - \max_{\Delta t_c, \Delta \phi_c}\frac{\langle h_1,h_2\rangle}{\sqrt{\langle h_1,h_1\rangle \langle h_2,h_2\rangle}}\,,
\end{equation}

where $\Delta t_c$ and $\Delta \phi_c$ are the difference in the time and phase
of coalescence, respectively, between the two waveforms. We randomly draw
source and lens parameters from the priors given in Table
\ref{tab:injection_priors}, generate the lensed waveforms using
\textsc{Gravelamps} and \textsc{Jim-lensing}, and compute the network mismatch
between them, which is the square root of the sum of the squared mismatches in
the different detectors. We use a network of LIGO-Livingston (L1), LIGO-Hanford
(H1) and Virgo (V1)~\citep{LIGOScientific:2014pky,VIRGO:2014yos}.  Throughout
this work, we use $f_\text{min} = 20\,$Hz, $f_\text{max} = 1024\,$Hz, the
\textsc{aLIGO\_O4\_high} PSD for the LIGO detectors and the \textsc{AdV} PSD
for Virgo~\citep{KAGRA:2013rdx,dcc_psds}.

\begin{table}
\caption{Priors for the injection study.}
\label{tab:injection_priors}
\begin{ruledtabular}
\begin{tabular}{lcc}
Parameter & Prior & Prior bounds \\
\hline
$\mathcal{M}/M_\odot$ & Uniform &$[20,50]$ \\
$q$ & Uniform &$[0.5,1.0]$ \\
$\chi_{1z},\chi_{2z}$ & Uniform &$[-0.8,0.8]$ \\
$d_L/\mathrm{Mpc}$ & $p(d_L)\propto d_L^2$ & $[100,1250]$ \\
$t_c/\mathrm{s}$ & Uniform &$[t_c-0.1,t_c+0.1]$ \\
$\phi_c$ & Uniform & $[0,2\pi]$ \\
$\iota$ & $p(\iota)\propto\sin\iota$ & $[0,\pi]$ \\
$\psi$ & Uniform & $[0,\pi]$ \\
$\alpha$ & Uniform & $[0,2\pi]$ \\
$\delta$ & $p(\delta)\propto\cos\delta$ & $[-\pi/2,\pi/2]$ \\
$M_{\rm L,z}/M_\odot$ & Uniform &$[10,5000]$ \\
$y$ & Uniform & $[0.1,3.0]$ \\
\end{tabular}
\end{ruledtabular}
\end{table}

\begin{figure}[t]
    \includegraphics[width=\linewidth]{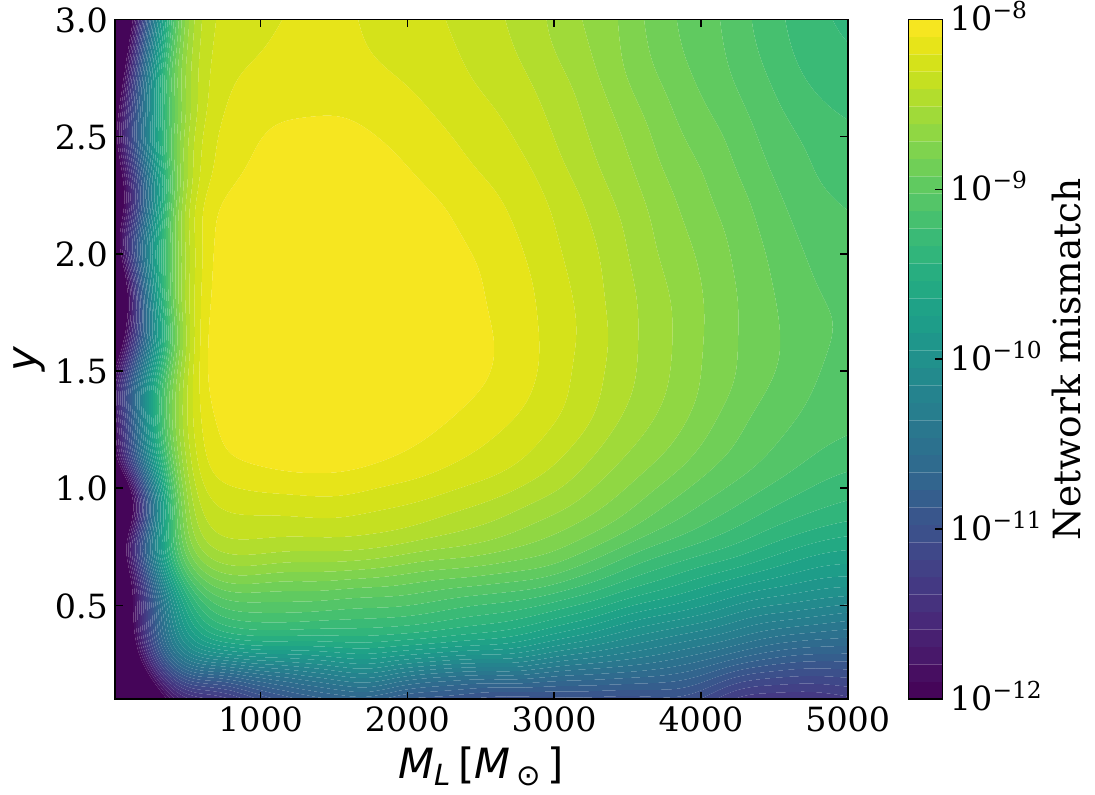}
    \caption{Network mismatch between lensed waveforms in \textsc{Gravelamps} and \textsc{Jim-lensing} for different lens parameters. The signal parameters are drawn randomly from the priors in Table \ref{tab:injection_priors}.}
    \label{fig:mismatch}
\end{figure}

As can be seen from Fig.  \ref{fig:mismatch}, the network mismatch does not
exceed $10^{-7}$, which is well below the maximum mismatch, $0.03$, in template
placements used by LVK matched filtering
searches~\citep{LIGOScientific:2025zdk} indicating strong consistency between
the two waveform construction approaches.

\subsection{Percentile-percentile plot}
\label{subsec:PP-plot}
We test if the PE in \textsc{Jim-lensing} is biased by constructing a
percentile-percentile (PP) plot, which is a standard test in data
analysis~\citep{Sidery_2014,Veitch:2014wba,Romero-Shaw:2020owr}. This entails
injecting a number of signals, whose parameters are drawn from a certain prior,
into noise and performing Bayesian analysis on the resulting data using the
same prior. For each injection $j$ and parameter $\theta_a$, we compute the
posterior percentile of the true value, \begin{equation}
r_{j,a}=\int_{-\infty}^{\theta^{\rm true}_{j,a}}
p(\theta_a|d_j)\,\dd\theta_a\,, \end{equation} where $p(\theta_a|d_j)$ is the
marginalized posterior distribution. For an unbiased recovery, the ranks
$r_{j,a}$ are uniformly distributed. When we plot the empirical cumulative
distribution for each parameter, which is referred to as a PP-plot, we should
obtain curves close to the diagonal. 

\begin{figure}[t]
    \includegraphics[width=\linewidth]{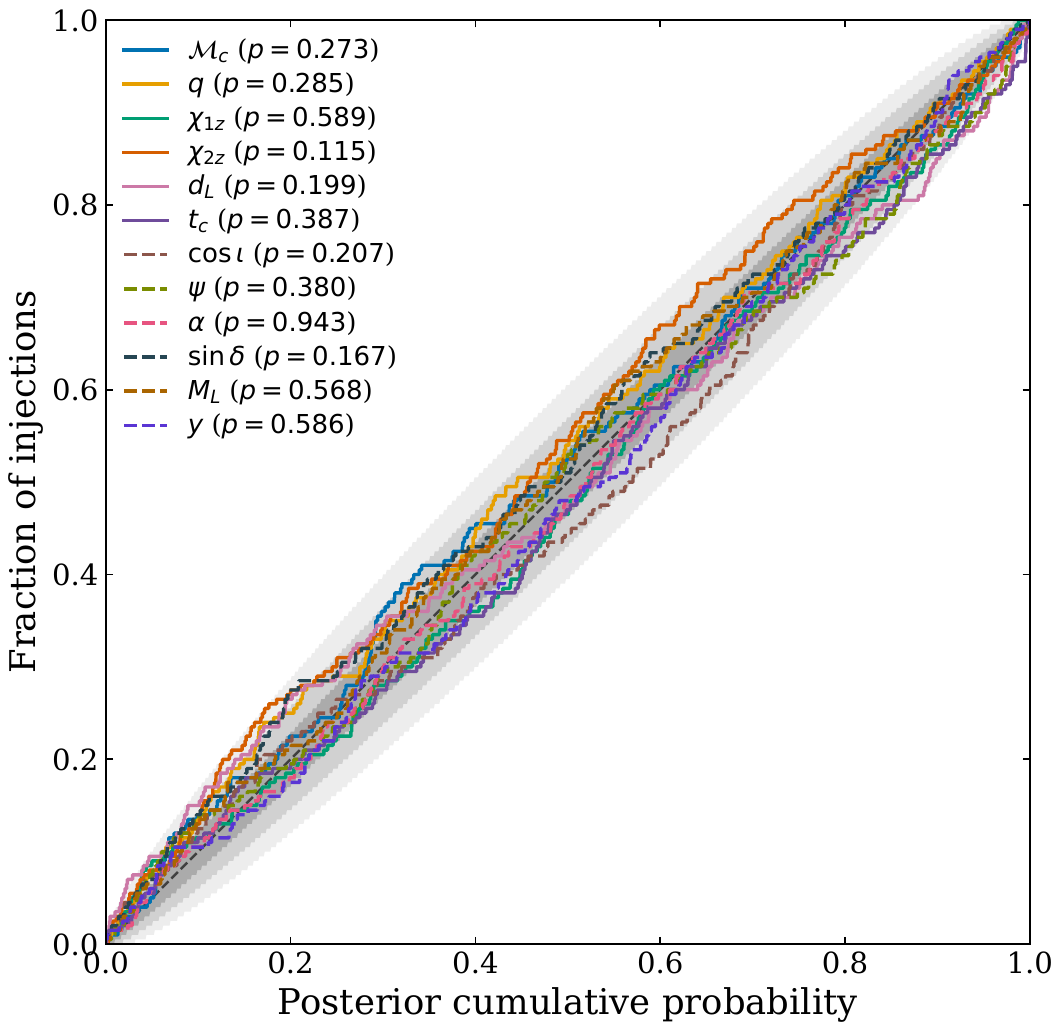}
    \caption{PP-plot for $200$ simulated signals injected in Gaussian noise. The values next to the parameter symbols are the $p$-values of a uniformity test using the KS statistic. The gray bands correspond to the $1$, $2$ and $3\sigma$ credible intervals.}
    \label{fig:PP_plot}
\end{figure}

We construct a PP-plot using $200$ injections drawn from the prior in Table
\ref{tab:injection_priors}. We use a network
of H1-L1-V1, resulting in frames of $4$ seconds sampled at $2048\,$Hz. For each
parameter, we also compute the Kolmogorov-Smirnov (KS) statistic, quantifying
how closely the ranks follow a uniform distribution. The resulting PP-plot,
including the $p$-values of the KS test, is shown in Fig. \ref{fig:PP_plot}.

We see that the curves trend along the diagonal. The combined $p$-value is $0.32$, showing that this method is unbiased. We also time the PP-plot runs and find that the median wall-clock duration is $17$ minutes when run on a NVIDIA A100 40/80GB.

\subsection{Posterior comparison}

\begin{figure*}[htbp] 
    \centering 
    \includegraphics[width=\textwidth]{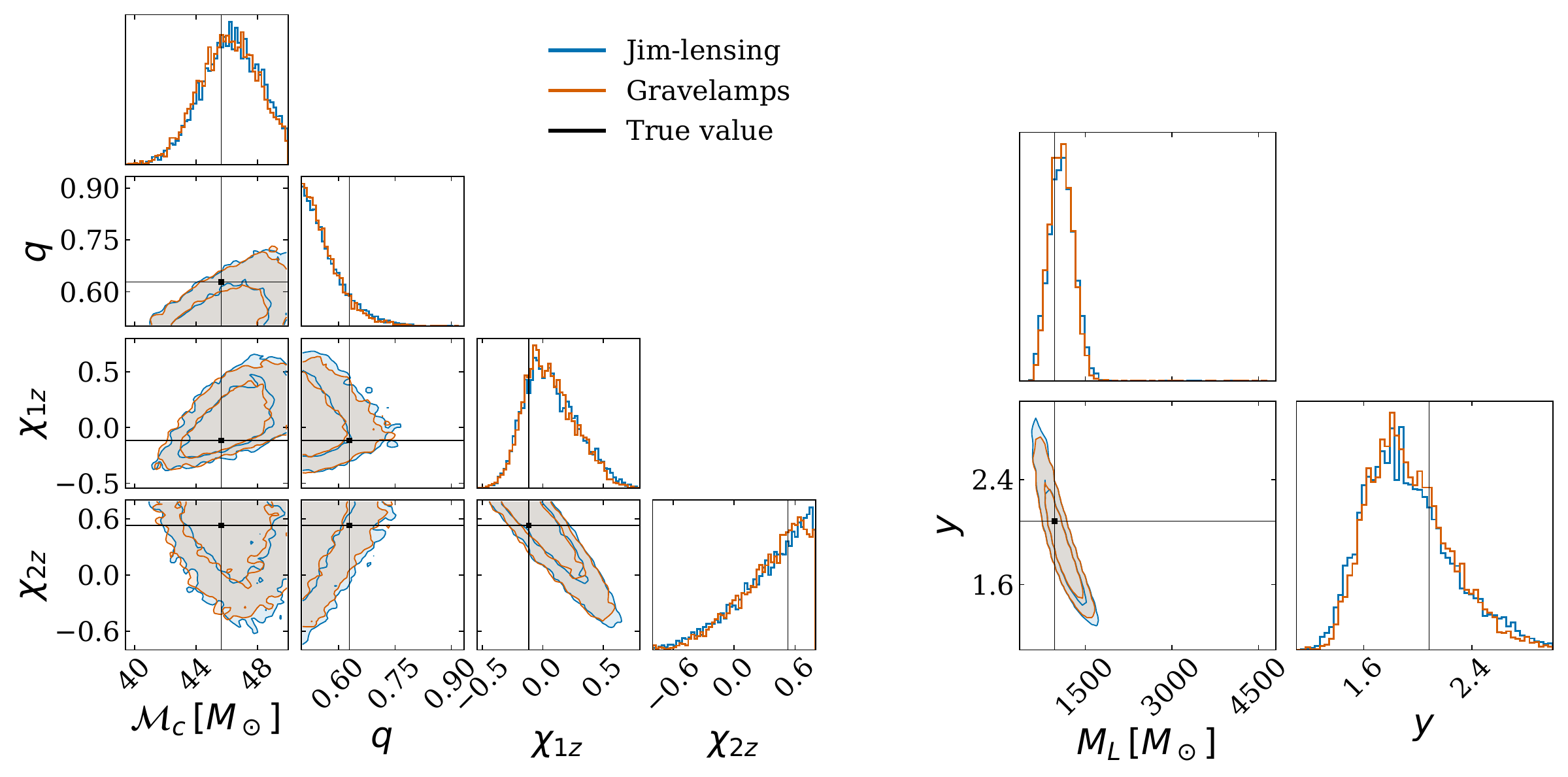}
    \caption{Corner plot of marginalized posterior distributions of the intrinsic and lens parameters of a reference run analyzed with \textsc{Jim-lensing} and \textsc{Gravelamps}. The extrinsic parameters are shown in Fig.~\ref{fig:corner_plot_extrinsic}.} 
    \label{fig:corner_plot} 
\end{figure*}

For $100$ signals from the PP-plot ensemble, we also perform PE using
\textsc{Gravelamps} and compare with \textsc{Jim-lensing}. An example of such a
comparison is shown in Fig. \ref{fig:corner_plot}. The median wall-clock duration for these runs is $16.3$ hours on $16$ CPUs, which is a factor of about $58$ slower than \jimlensing{}.

\begin{figure}[t]
    \includegraphics[width=\linewidth]{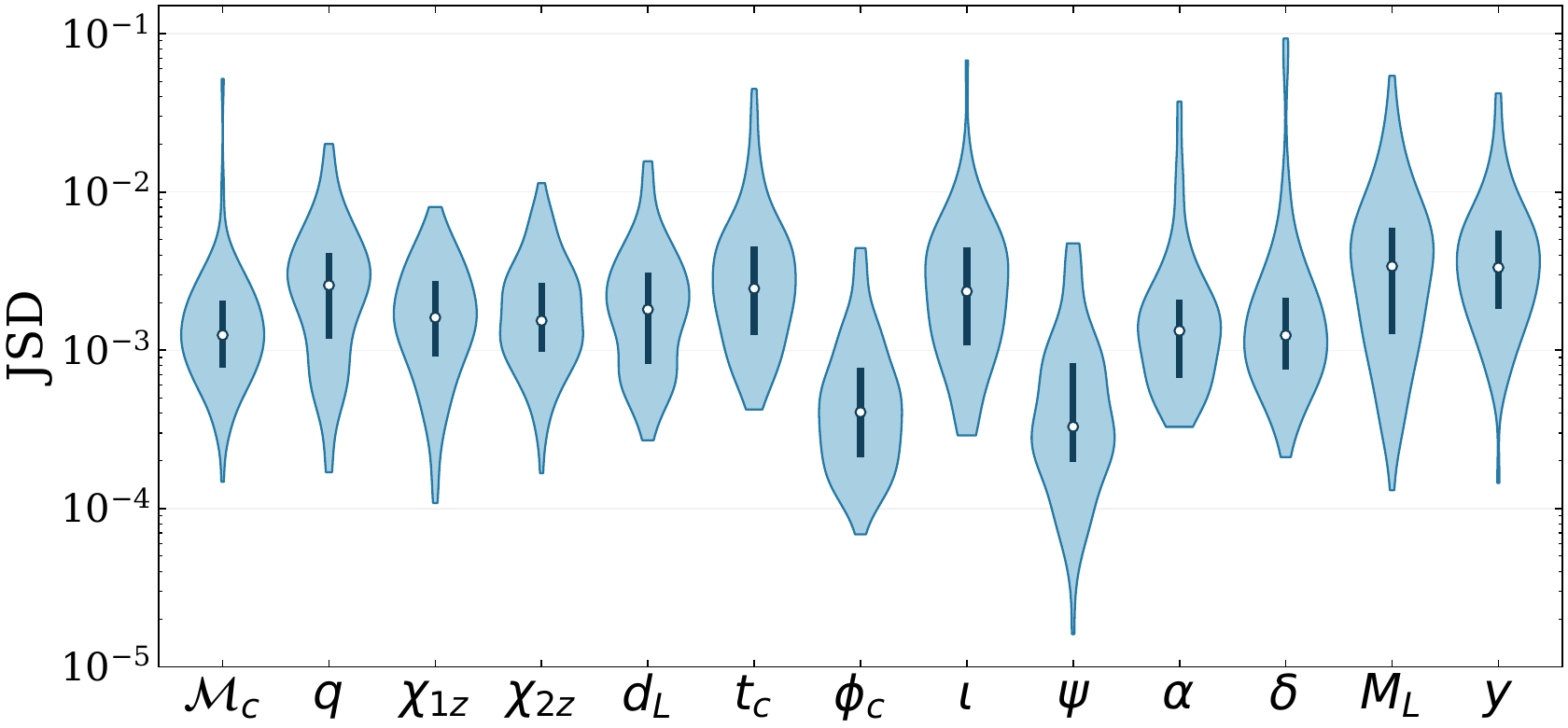}
    \caption{Violin plots of JSD values comparing \jimlensing{} and \textsc{Gravelamps} for all marginalized single-parameter posterior distributions. The vertical bars show the $25$-$75$\% confidence intervals.}
    \label{fig:JSD_hist}
\end{figure}

To quantify the difference between the posterior distributions given by
\textsc{Jim-lensing} and \textsc{Gravelamps}, we compute for each marginalized
posterior distribution the Jensen-Shannon divergence (JSD), which is a measure
of distance between two probability distributions. The JSD is zero for
identical distributions and $\ln 2$ for maximally divergent distributions. In
Fig.~\ref{fig:JSD_hist}, the distribution of JSD values for all marginalized single-parameter
posterior distributions are shown as violin plots.

We see that the bulk of the JSD values are below the $2.0\times 10^{-3}$
threshold which is considered by~\citet{Romero-Shaw:2020owr} to be
statistically insignificant run-to-run variance. Of those analyses that lie
above this threshold, most of the values remain below the $1.7\times 10^{-2}$ maximal JSD
found by~\citet{Wong:2023lgb} in the \textsc{Jim} reanalysis of GW150914, with only a few exceptions, although they are still bounded by $9.4\times 10^{-2}$. We show the $10$ posteriors with the highest JSD value in Appendix \ref{subsec:corner_plot_extrinsic} in Fig.~\ref{fig:highest_JSD}, demonstrating that even for the largest divergences, the posterior distributions remain comparable. Altogether, accounting for differences in sampling methodology implementation,
this indicates considerable agreement between \textsc{Gravelamps} and
\textsc{Jim-Lensing}.

\subsection{Bayes factor recovery}
To further validate our implementation, we compute the Bayes factor $\mathcal{B}^\mathrm{ML}_\mathrm{UL}$ for $200$ unlensed and $200$ lensed injections in Gaussian noise. The source parameters are drawn from an astrophysical prior using \textsc{LeR}~\citep{Phurailatpam:2024enk}, a software package for simulating unlensed and strongly lensed CBC populations. The lensing parameters are drawn from the priors given in Table~\ref{tab:injection_priors}, ensuring, for the lensed injections, that the mismatch between the lensed and unlensed waveform is at least $0.05$. 
Furthermore, we use likelihood reweighting~\citep{Payne:2019wmy} by \textsc{Bilby}, for the \jim{} PE runs, and by \textsc{Gravelamps}, for the \jimlensing{} PE runs, to compare the convergence of our method with state-of-the-art pipelines that are used to compute point-mass lens Bayes factors in the LVK~\citep{LIGOScientific:2023bwz,LIGOScientific:2025cwb}. The result of this reweighting is shown in Appendix~\ref{subsec:rew_BF}, demonstrating that the reweighting efficiency remains high across the entire injection set, with a minimum of $99.43$\%. This shows that our framework is able to accurately obtain Bayes factors, avoiding the potential issues associated with out-of-training data, which can impede accurate Bayes factor computation in accelerated lensing pipelines that rely on trained models~\citep{Caldarola:2025oxr}.

\begin{figure}[t]
    \includegraphics[width=\linewidth]{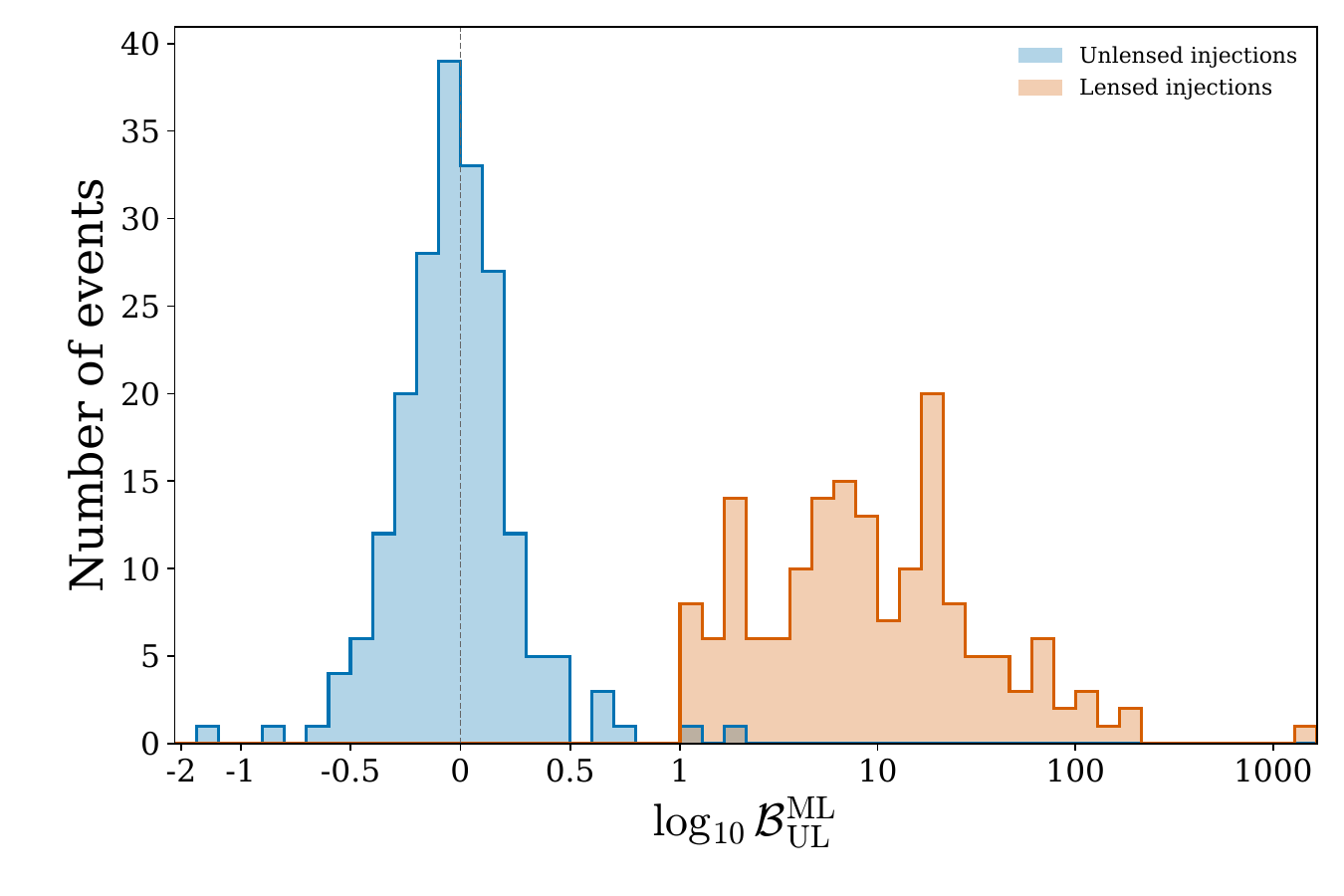}
    \caption{Distribution of log$10$ Bayes factors of a background of $200$ events and a foreground of $165$ events obtained by \jimlensing{}.}
    \label{fig:BF_hist}
\end{figure}

We apply a filter of the reweighted \textsc{Gravelamps} and \textsc{Bilby} log$10$ Bayes factor greater than $1$ for a lensed event to be detectable, which is a standard threshold used in LVK analyses~\citep{LIGOScientific:2025cwb}, retaining $165$ events in the foreground. The result of this study is shown in Fig.~\ref{fig:BF_hist}.
For the unlensed background, we obtain similar numbers to the background analyzed by \textsc{Gravelamps} in \cite{LIGOScientific:2025cwb}. For the lensed foreground, we obtain log$10$ Bayes factors greater than $1$, meaning that all retained lensed events are detectable by \jimlensing{}, as expected by the high reweighting efficiencies. We conclude that \jimlensing{} can be used as a rapid and robust method to compute lensing Bayes factors and hence identify lensed GW signals.

\section{Conclusion}
\label{sec:conclusion}
In this work, we present \jimlensing{}, a GPU-accelerated Bayesian
inference framework for wave-optically lensed GW signals. This is the first
such approach that can perform accurate Bayes factor estimates, without data training requirements. We achieve this by
integrating new methods for efficient computation of the point mass
amplification factor into the \textsc{JAX} ecosystem and integrating it within
the \textsc{Jim} PE pipeline. This enables accurate inference, using similar
methods to the current state-of-the-art meaning deployment is possible across
all network configurations and scales, whilst retaining the significant
performance benefits of GPU-acceleration. These performance gains allow PE
to be performed in the order of $10$ minutes, compared to the hours-to-days of current CPU
infrastructure. Furthermore, this approach can readily be augmented to include
additional lens models in the future.

We validated our implementation through several tests. First, a waveform-level
comparison against \textsc{Gravelamps} demonstrated excellent agreement, with
network mismatches well below levels expected to impact posterior recoveries.
Second, a PP-test based on $200$ simulated injections showed that the recovered
posterior distributions are statistically consistent with unbiased inference.
Finally, direct comparisons of posterior distributions obtained with
\jimlensing{} and \textsc{Gravelamps} show strong agreement across all
inferred parameters.

Forthcoming implementations of modern waveform models in \textsc{ripple} will
enable \jimlensing{} to reach complete parity with the current
\textsc{bilby}-based infrastructure for lensed GW data analysis in waveform
physics. The significantly increased performance will then alleviate the
problems associated with timely analysis of an ever increasing rate of GW
detections. Furthermore, it will also allow the construction of much deeper
backgrounds of simulated signals to allow frequentist assessments of
significance of any potential candidates. 

Future work on \jimlensing{} will focus on augmenting
analyses to more complex lens and waveform models broadening the range of
possible lensing scenarios that may be robustly detected, increasing the
chances of correctly identifying any future candidates and unlocking the vast
array of scientific avenues of study that gravitational lensing of GW signals
promises.

\section*{Acknowledgements}
We thank Thibeau Wouters for the valuable input regarding \jim{}. We thank Jose Ezquiaga for the careful review. Z. L. is partly supported by the National Key Research and Development Program of China (No.~2023YFC2206704), and the Fundamental Research Funds for the Central Universities. J. H. is a FRIA grantee of the Fonds de la Recherche Scientifique - FNRS. M.W., T.C.K.N. and C.V.D.B acknowledge support by the research programme of the Netherlands Organisation for Scientific Research (NWO). J. J. acknowledges support from the Fonds de la Recherche Scientifique and the Royal Observatory of Belgium. Computational resources have been provided by the Consortium des \'{E}quipments (C\'{E}CI) funded by the Fonds de la Recherche Scientifique de Belgique (F.R.S.-FNRS) under Grant No. 2.4020.11 and by the Walloon Region.
\appendix
\label{sec:appendix}
\section{Confluent hypergeometric function}
\label{subsec:1F1}
To ease notation, we define
\begin{align}
H_b={}_1F_1(iu;b;i\zeta),\qquad
\end{align}
so that the required function is $H_1$.
We use four kinds of approximations, which we call a power series, a downward recurrence, a large-argument expansion, and a Temme-type Bessel expansion.

\paragraph{Power series}
The defining power series~\cite[Eq.~13.2.2]{DLMF} gives
\begin{equation}
H_b=\sum_{n=0}^{\infty}t_n^{(b)},\qquad
t_n^{(b)}=\frac{(iu)_n}{(b)_n}\frac{(i\zeta)^n}{n!},
\label{eq:sf_kummer_series}
\end{equation}
where the rising factorials are
\begin{align*}
(iu)_n&=\prod_{k=0}^{n-1}(iu+k),\qquad\\
(b)_n&=\prod_{k=0}^{n-1}(b+k),\qquad\\
(iu)_0&=(b)_0=1.
\end{align*}
Starting from $t_0^{(b)}=1$, successive terms are generated by
\begin{equation}
t_n^{(b)}=t_{n-1}^{(b)}
\frac{-u\zeta+i(n-1)\zeta}{n(n+b-1)},\qquad n\geq1.
\label{eq:sf_series_terms}
\end{equation}
Setting $b=1$ gives the required function $H_1$.

\paragraph{Downward recurrence}
Instead of evaluating the series directly at $b=1$, we first compute
$H_{b_{\rm start}+1}$ and $H_{b_{\rm start}+2}$ from the same power series,
where $b_{\rm start}$ is a positive integer specifying the starting index.
We then apply the contiguous relation~\cite[Eq.~13.3.2]{DLMF},
\begin{equation}
H_b=\left(1+\frac{i\zeta}{b}\right)H_{b+1}
-\left(1-\frac{iu}{b+1}\right)\frac{i\zeta}{b}H_{b+2},
\label{eq:sf_downward}
\end{equation}
successively for $b=b_{\rm start},b_{\rm start}-1,\ldots,1$ to obtain $H_1$.

\paragraph{Large-argument expansion}
We use the two-component asymptotic expansion of the Kummer function given in
Ref.~\cite[Eq.~13.7.2]{DLMF}. Setting $a=iu$, $b=1$, and $z=i\zeta$ on the
positive imaginary axis, the expansion for fixed $a$ as $|z|\to\infty$ is
\begin{equation}
H_1\sim
\frac{e^z z^{a-1}}{\Gamma(a)}
\sum_{k=0}^{\infty}\frac{(1-a)_k^2}{k!z^k}
+\frac{e^{i\pi a}z^{-a}}{\Gamma(1-a)}
\sum_{k=0}^{\infty}\frac{(a)_k^2}{k!(-z)^k}.
\label{eq:sf_largez}
\end{equation}
Here the positive sign in $e^{i\pi a}$ corresponds to $z=i\zeta$,
$\zeta>0$, and complex powers use the principal logarithm.

\paragraph{Temme-type Bessel expansion}
We use Temme's Bessel expansion~\cite[Eq.~27.4.63]{Temme2014},
with $\lambda=0$, to obtain
\begin{equation}
H_1\sim e^{iuc/2}
\left[
J_0(x)\sum_{k=0}^{\infty}\frac{a_k(c)}{(iu)^k}
-i\beta J_1(x)\sum_{k=0}^{\infty}\frac{b_k(c)}{(iu)^k}
\right].
\label{eq:sf_temme}
\end{equation}
where
\begin{equation}
c=y^2,\quad\eta=2\operatorname{asinh}(\sqrt{c}/2),\quad x=u(\eta+\sinh\eta).
\end{equation}
The coefficients are defined by
\begin{align}
a_k(c)&=\frac{f_k(\beta;c)+f_k(-\beta;c)}{2},
\qquad\\
b_k(c)&=\frac{f_k(\beta;c)-f_k(-\beta;c)}{2\beta},
\end{align}
where $f_k$ follows Temme's recurrence
\cite[Eq.~(27.1.9)]{Temme2014}, initialized by
Eq.~(27.4.39) with $\lambda=0$.

\paragraph{Bessel functions}

For the Bessel functions in \eqref{eq:sf_temme}, we use for fixed $\nu$ and $x\to+\infty$, the Hankel
expansion~\cite[Eqs.~10.17.1--10.17.3]{DLMF} is
\begin{align}
J_\nu(x)\sim\sqrt{\frac{2}{\pi x}}
\{
&\cos\chi_\nu\sum_{k=0}^{\infty}\frac{(-1)^k C_{2k}(\nu)}{x^{2k}}\\
&-\sin\chi_\nu\sum_{k=0}^{\infty}\frac{(-1)^k C_{2k+1}(\nu)}{x^{2k+1}}
\},
\end{align}
where
\begin{align}
\chi_\nu&=x-\frac{\nu\pi}{2}-\frac{\pi}{4},\qquad
C_0(\nu)=1,\qquad\\
C_m(\nu)&=\frac{\prod_{j=1}^{m}[4\nu^2-(2j-1)^2]}{m!\,8^m}.
\end{align}
Setting $\nu=0$ or $1$ gives the expansions for $J_0$ and $J_1$.

\section{Posterior comparison}
\label{subsec:corner_plot_extrinsic}
\begin{figure*}[]
    \centering
    \includegraphics[width=\linewidth]{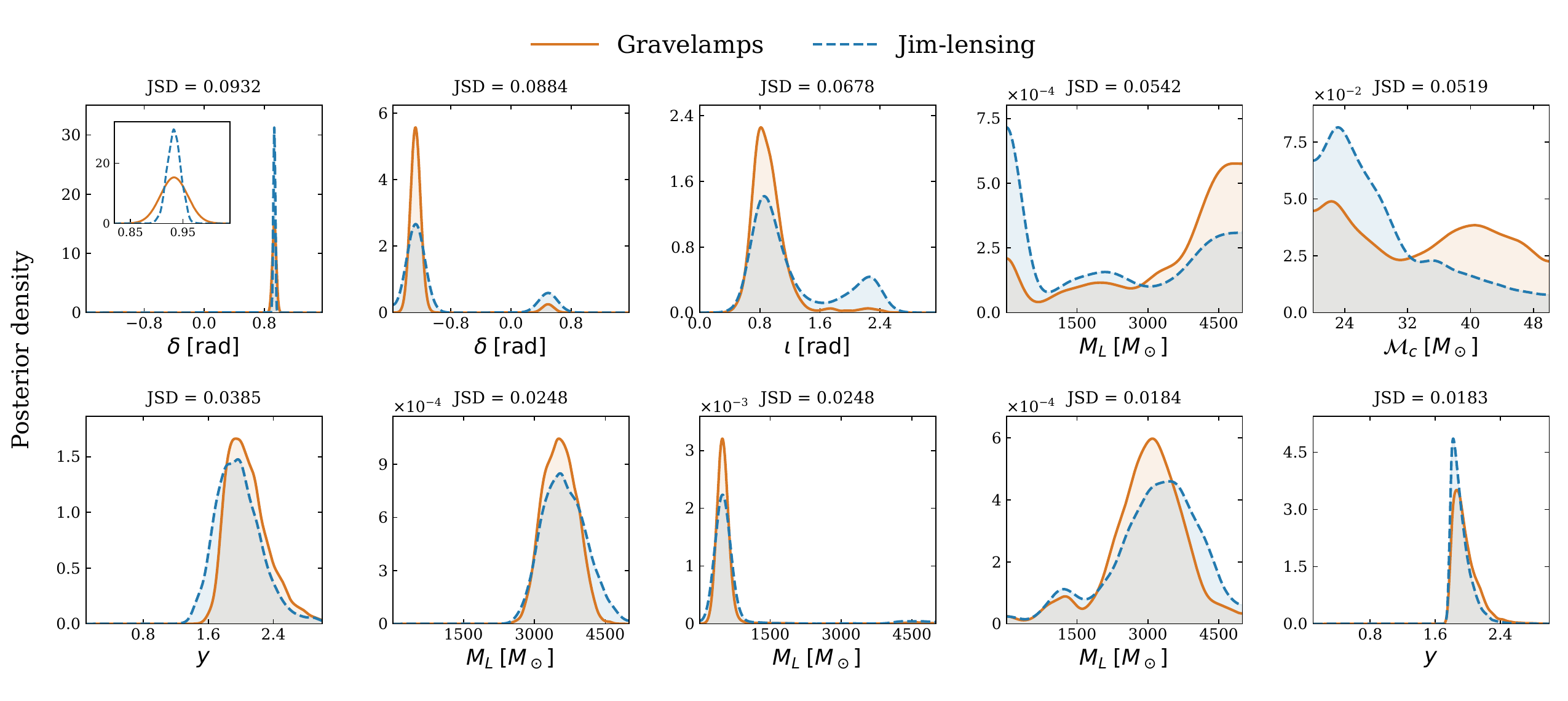}
    \caption{Posterior distributions of the cases with the highest JSD values between \jimlensing{} and \textsc{Gravelamps}.}
    \label{fig:highest_JSD}
\end{figure*}

\begin{figure*}
\begin{minipage}[t]{0.48\textwidth}
    \centering
    \includegraphics[width=0.9\columnwidth]{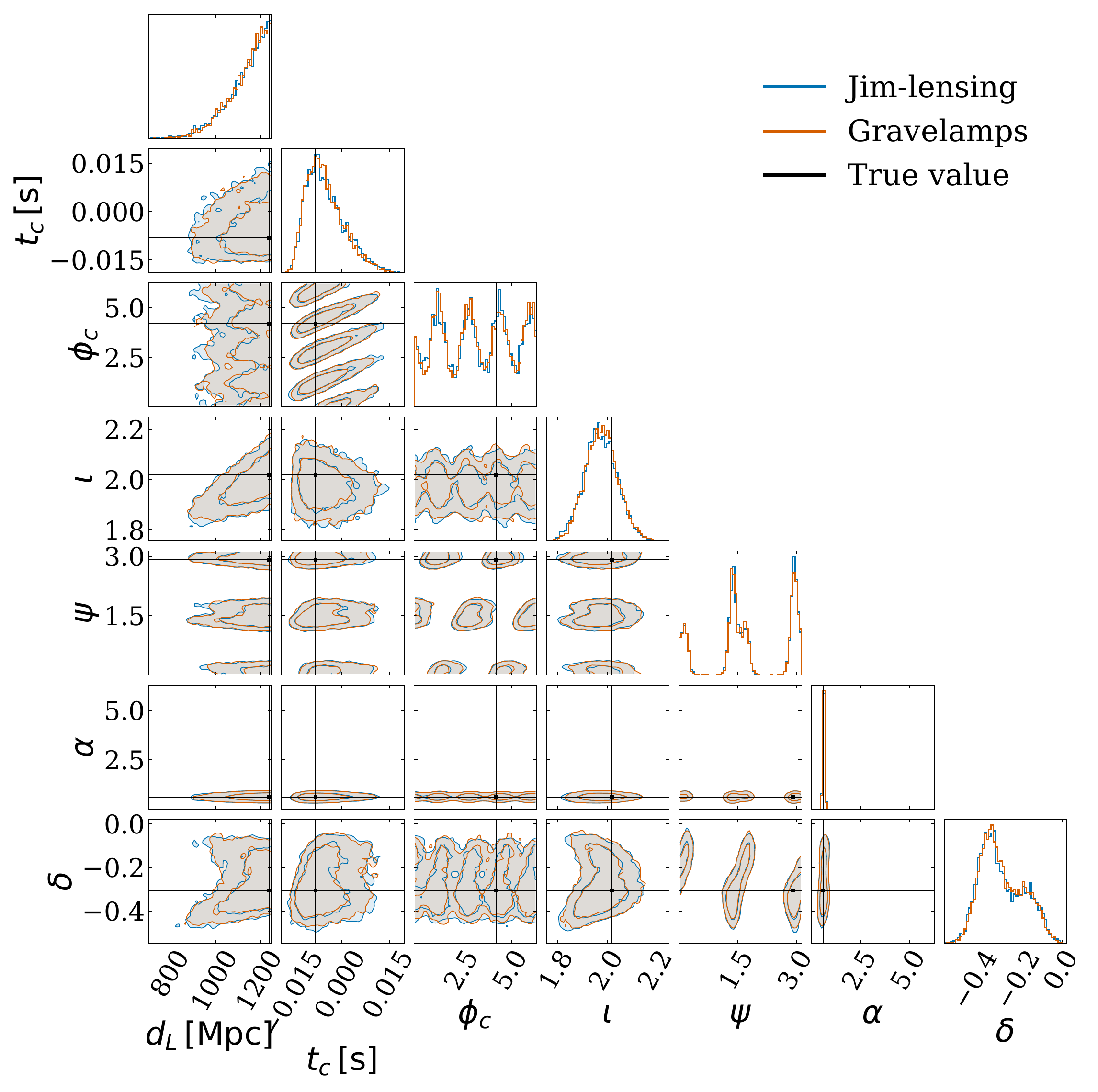}
    \caption{Comparison between the marginalized posteriors of the extrinsic parameters of a reference run. The coalescence phase posterior of the \jimlensing{} run is reconstructed in post-processing.}
    \label{fig:corner_plot_extrinsic}
\end{minipage}
\hfill
\begin{minipage}[t]{0.48\textwidth}
    \centering
    \includegraphics[width=0.9\columnwidth]{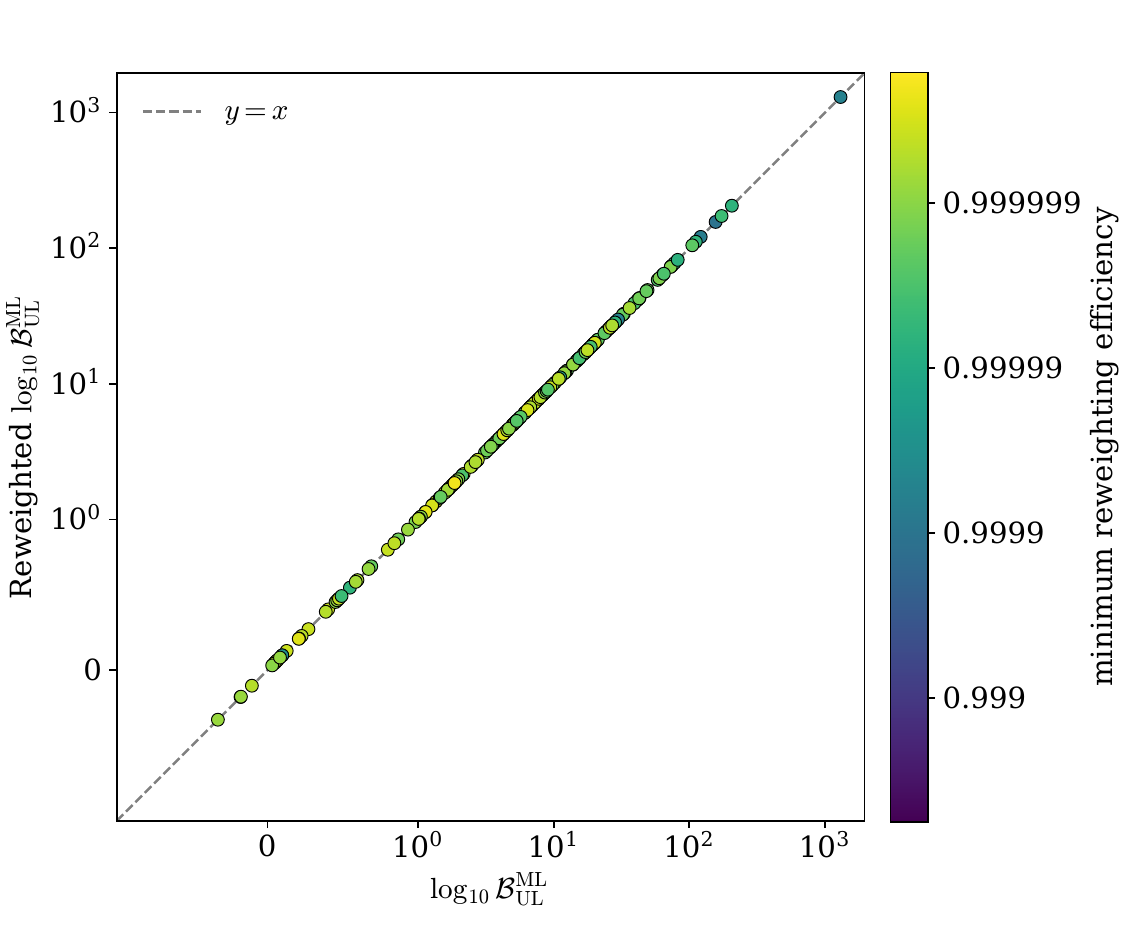}
    \caption{Bayes factors obtained using \jim{} and \jimlensing{} compared with the reweighted values using \textsc{Bilby} and \textsc{Gravelamps}. The colorbar shows the minimum reweighting efficiency between unlensed and lensed PE runs. The minimum reweighting efficiency across the entire set is $99.43$\%.}
    \label{fig:BF_vs_BF}
\end{minipage}
\end{figure*}

We show a comparison between the marginalized posteriors of the extrinsic parameters of a reference run from the PP-plot ensemble from Section \ref{subsec:PP-plot} in Fig.~\ref{fig:corner_plot_extrinsic}, demonstrating good agreement between \jimlensing{} and \textsc{Gravelamps}. Furthermore, we show in Fig.~\ref{fig:highest_JSD} the $10$ posterior distributions with the highest JSD values between \jimlensing{} and \textsc{Gravelamps}.
\section{Reweighted Bayes factors}
\label{subsec:rew_BF}
We reweight the evidences of the lensed injections obtained using \jim{} and \jimlensing{}, by \textsc{Bilby} and \textsc{Gravelamps}, respectively, to obtain the reweighted Bayes factor.
As can be seen in Fig.~\ref{fig:BF_vs_BF}, the reweighting efficiency is high across injections, both for the lensed and unlensed populations, with the minimum reweighting efficiency being $99.43$\%.
\bibliographystyle{apsrev4-2}
\bibliography{references}

\end{document}